\documentclass[prl,reprint,twocolumn,superscriptaddress,showpacs,floatfix]{revtex4-1}
\usepackage{amsmath}
\usepackage{mathrsfs}
\usepackage{txfonts}
\usepackage{amssymb}
\usepackage{graphicx}
\usepackage{hyperref}
\usepackage{ulem}
 \usepackage{overpic}
 \usepackage{psfrag}
\usepackage{tabularx}
\usepackage{array}
\usepackage{placeins}
\newcommand{\PreserveBackslash}[1]{\let\temp=\\#1\let\\=\temp}

\makeatletter

\begin{document}

\title{Entanglement entropy for scale-invariant states: universal finite-size scaling}

\author{Huan-Qiang Zhou}
\affiliation{Centre for Modern Physics, Chongqing University, Chongqing 400044, The People's Republic of China}

\author{Qian-Qian Shi}
\affiliation{Centre for Modern Physics, Chongqing University, Chongqing 400044, The People's Republic of China}

\author{Ian P. McCulloch}
\affiliation{School of Mathematics and Physics, The University of Queensland, St. Lucia, QLD 4072, Australia}

\author{Murray T. Batchelor}
\affiliation{Centre for Modern Physics, Chongqing University, Chongqing 400044, The People's Republic of China}
\affiliation{Mathematical Sciences Institute, The Australian National University, Canberra ACT 2601, Australia}

\begin{abstract}
A universal finite system-size scaling analysis of the entanglement entropy is presented for highly degenerate ground states arising from spontaneous symmetry breaking with type-B Goldstone modes in exactly solvable one-dimensional quantum many-body systems. These states appear to be scale-invariant, but not conformally invariant. Our findings are based on a physical argument, imposing three constraints  on the entanglement entropy, in addition to further confirmation from an asymptotic analysis of
the entanglement entropy for the ${\rm SU}(2)$ spin-$1/2$ ferromagnetic states. The resulting universal scaling form is demonstrated for three fundamental models -- the ${\rm SU}(2)$ spin-$s$ Heisenberg ferromagnetic model, the ${\rm SU}(N+1)$ ferromagnetic model, and the staggered ${\rm SU}(3)$ spin-1 ferromagnetic  biquadratic model. The results point towards a classification for distinct types of scale-invariant states,  relevant to a complete classification of quantum states of matter.
\end{abstract}.
\maketitle

\section{Introduction}

Symmetry offers a powerful means to understand fundamental phenomena in physics and beyond~\cite{symmetry,andersonbook,SSBbook}. A specific example concerns scale invariance, which was believed to imply conformal invariance,  as first speculated by Polyakov~\cite{polyakov}.  Historically, this speculation led to the creation of conformal field theory~\cite{cft}. Since then much effort has been made in an attempt to clarify a deep connection between scale and conformal invariance~\cite{morecf}.
However, a systematic investigation of scale-invariant, but not conformally invariant, quantum states, relevant to a classification of quantum critical points, is still lacking, even for one-dimensional quantum many-body systems.

A recent conceptual development has revealed that highly degenerate ground states arising from spontaneous symmetry breaking with type-B Goldstone modes in one-dimensional quantum many-body systems are scale-invariant~\cite{FMGM,golden,LLspin1}.
Somewhat surprisingly, a paradigmatic example is the ${\rm SU}(2)$ Heisenberg ferromagnet, with the highly degenerate ground states being the familiar ferromagnetic states.
As demonstrated,  the highly degenerate ground states admit an exact singular value decomposition, thus exhibiting self-similarities in the ground state subspace.
As a consequence, an abstract fractal is revealed, living in a Hilbert space, which may be characterized in terms of the fractal dimension,
introduced earlier by Castro-Alvaredo and Doyon~\cite{doyon} for the ${\rm SU}(2)$ Heisenberg ferromagnetic states in the context of a field-theoretic approach.
The fractal dimension is identified with the number of the type-B Goldstone modes~\cite{FMGM}. In other words, there is a fascinating connection between scale-invariant states and the counting rule of the Goldstone modes~\cite{FMGM}. The latter in turn is established with the introduction of the Watanabe-Brauner matrix~\cite{brauner-watanabe,watanabe}, as a result of an insightful observation made by Nambu~\cite{nambu}, thus leading to the classification of type-A and type-B Goldstone modes.

However, a natural question arises as to whether or not it is possible to distinguish scale-invariant states from conformally invariant states,
given that the entanglement entropy scales logarithmically with the block size in the thermodynamic limit, with the prefactor being half the number of the type-B Goldstone modes~\cite{FMGM} for scale-invariant states and $c/3$ for conformally invariant states, where $c$ is the central charge.
One may anticipate that a finite system-size scaling behavior of the entanglement entropy is different for scale-invariant states and conformally invariant states, thus furnishing a sensible means for distinguishing them. Physically, such a distinct finite system-size scaling behavior is deeply rooted in the fact that highly degenerate ground states arising from spontaneous symmetry breaking with type-B Goldstone modes {\it only} concerns a scale transformation, instead of being a scale transformation plus the Lorentz boost for conformally invariant states.

This question is addressed here using an heuristic physical argument, which leads to three constraints imposed on the entanglement entropy: first, it must be identical for a block and for its environment;  second, it should reproduce a logarithmic scaling function with the block size in the thermodynamic limit; third, as a function of both the system size and the block size,
it must be homogeneous, with the order being one, to render consistency with a scale transformation. As it turns out, the universal scaling function for such a type of scale-invariant states is the
most fundamental solution satisfying the three constraints, with further confirmation coming from an asymptotic analysis of
the entanglement entropy for the ${\rm SU}(2)$ spin-$1/2$ ferromagnetic states.
To illustrate our scheme, we choose the ${\rm SU}(2)$ spin-$s$ Heisenberg ferromagnetic model, the ${\rm SU}(N+1)$ ferromagnetic model, and the staggered ${\rm SU}(3)$ spin-1 ferromagnetic  biquadratic model.

\section{Generalities}
	
Consider a quantum many-body system on a lattice, described by Hamiltonian $\mathscr{H}$.
Suppose the system is partitioned into a block $B$ and its environment $E$, with the block consisting of $n$ (contiguous) lattice sites,
with the other $L-n$ lattice sites constituting the environment $E$.
If the symmetry group $G$ is spontaneously broken into $H$,  then the counting rule  is established as $N_A+2N_B=N_{BG}$~\cite{watanabe},
where $N_A$ and $N_B$ are the numbers of the type-A and type-B GMs, and $N_{BG}$ is equal to the dimension of the coset space $G/H$.
In particular, for a quantum many-body system, no type-A GM survives in one spatial dimension, as follows from the Mermin-Wagner-Coleman theorem~\cite{mwc}.
Hence the number of the type-A GMs $N_A$ is always zero, if one is restricted to one-dimensional quantum many-body systems.

Here we present an heuristic argument to establish a universal scaling behavior of the entanglement entropy $S(L,n)$ with $L$ and $n$ for a scale-invariant quantum state, which is one of the highly degenerate ground states for a quantum many-body system undergoing spontaneous symmetry breaking with type-B Goldstone modes.
For completeness, an alternative derivation is also suggested for conformally invariant quantum states.

For both scale-invariant and conformally invariant states~\cite{FMGM}, the entanglement entropy $S(n)$  exhibits a logarithmic scaling behavior with the block size $n$ in the thermodynamic limit (up to an additive constant):
\begin{equation}
S(n) \propto \log_2 n.
\label{ln1}
\end{equation}
Here, $S(n)$ denotes $S(L,n)$, when $L \rightarrow \infty$. Note that the prefactor is  $N_B/2$ for a scale-invariant quantum state and $c/3$ for a conformally invariant quantum state, where $N_B$ denotes the number of the type-B Goldstone modes.
We remark that, physically, both central charge $c$ and the number of the type-B Goldstone modes $N_B$ count the number of gapless low-lying excitations for the quantum many-body system under investigation.

Keeping this fact in mind,  one may assume that, when the system size $L$ and the block size $n$ are finite but large enough, the entanglement entropy $S(L,n)$ scales with $L$ and $n$  as follows,
\begin{equation}
S(L,n)\propto \log_2 g(L,n). \label{ln2}\\
\end{equation}
Here, $g(L,n)$ is introduced as a universal scaling function of $L$ and $n$, yet to be determined.
Three constraints imposed on $g(L,n)$ emerge
from the following three physical requirements, to which the entanglement entropy $S(L,n)$ is subject.

First, the fact that for any pure quantum state, the entanglement entropy for a block $B$ is identical to that for its environment $E$~\cite{nielsen} implies that
$S(L,n)= S(L,L-n)$. Hence, we have
\begin{equation}
g(L,n)= g(L,L-n). \label{con1}
\end{equation}
In addition, $S(L,n)$ must be monotonically increasing with $n$ until it reaches $L/2$, where $L$ is assumed to be even.

Second, as the thermodynamic limit is approached, i.e.,  $L\rightarrow \infty$, Eq.~(\ref{ln2}) reduces to Eq.~(\ref{ln1}).
Therefore, one may expect that in this limit $g(L,n)$ becomes $n$:
\begin{equation}
\lim_{L\rightarrow\infty} g(L,n) =  n. \label{con2}
\end{equation}
Third, $g(L,n)$ must be a first-order homogeneous function of $L$ and $n$ as a result of a scale transformation: $n \rightarrow \lambda \, n$ and $L\rightarrow \lambda \, L$, given that it becomes $n$ in the thermodynamic limit [cf. Eq.~(\ref{con2})]. Thus
\begin{equation}
g(\lambda  L,\lambda n)=\lambda \, g(L,n).
\end{equation}

As it turns out, the three constraints imposed on $g(L,n)$ allow us to determine its specific forms.
In fact, if one chooses $\lambda=1/L$,  then we have
\begin{equation}
g(L,n)=L \, g(1,n/L).
\end{equation}
This implies that $g(L,n)$ is, up to a multiplicative factor $L$, a function of $n/L$:
\begin{equation}
g(L,n)=L \, k(n/L).
\end{equation}
Here, $k(n/L)$ is a function of $n/L$, yet to be determined. Our task is therefore reduced to the determination of the function $k(x)$, with $x=n/L$.

To proceed, let us first make an intriguing observation.
Suppose that a function $k(x)$ yields a solution satisfying the three constraints, then any function $F(k(x))$ of $k(x)$ also yields a solution, subject to two conditions for $F(x)$: (i) its constant term vanishes,  $F(0)=0$ and (ii) its first-order derivative is nonzero, $F'(0) \neq 0$.
This observation drastically simplifies our task, since we may restrict ourselves to searching for the most fundamental solution, corresponding to a plain scale transformation, in the sense that no other symmetry transformation is involved, if one excludes any internal symmetry.

For our purpose, it is plausible to assume that $k(x)=u(x)u(1-x)$, with $u(x)$ being a polynomial function of $x$, in order to ensure that the first constraint (\ref{con1}) is satisfied.
Then, the second constraint (\ref{con2}) implies that $u(x)=x$ as the simplest choice. Hence, we have $k(x)=x(1-x)$. This yields the simplest universal scaling function, thus constituting the most fundamental solution to the three constraints for scale-invariant states.
Here, we remark that there exists another solution $k_c(x)$, which may be expressed in terms of elementary functions, if one assumes that $k_c(x)=v(x)+v(1-x)$, with $v(x)$ being a
non-polynomial function of $x$,  in order to ensure that the first constraint (\ref{con1}) is satisfied. Then, the second constraint (\ref{con2}) implies that $v(x)= 1/(2i\pi) \exp(i\pi x)$. Hence, we have $k_c(x)=1/\pi\sin (\pi x)$.
Indeed, $k_c(x)$ may be re-expressed as a function of $k(x)$, namely $k_c(x) = 1/\pi \cos [\pi/2\sqrt{1-4k(x)}]$, as one may have anticipated from the observation above.
Actually,  $k_c(x)$ constitutes a universal scaling function for conformally invariant
states, as predicted by Calabrese and Cardy~\cite{cardy} (more details on the universal scaling functions $k(x)$ and $k_c(x)$ are given in Section A of the Supplemental Material (SM)).

In order to lend further support to our claim that $k(x)$ is the desired universal scaling function for scale-invariant states arising from spontaneous symmetry breaking with type-B Goldstone modes, we may resort to exactly solvable quantum many-body systems.
One may choose the ${\rm SU}(2)$ spin-$1/2$ ferromagnetic Heisenberg model, with the number of the type-B Goldstone mode $N_B$ being 1.
For this model, the entanglement entropy $S(L,n)$ has been derived by Popkov and  Salerno~\cite{popkov}.
As it turns out, our findings reproduce their results for the ${\rm SU}(2)$ spin-$1/2$ ferromagnetic states (cf. Section B of the SM).
For a conformally invariant state, one may choose the transverse-field spin-$1/2$ Ising chain at its critical point, with central charge $c=1/2$.
For this model, the entanglement entropy $S(L,n)$ has been derived by Jin and Korepin~\cite{korepin}, thus allowing us to reproduce the prediction from conformal field theory~\cite{cardy}.

More precisely, for a scale-invariant state, with a given filling $f$, the entanglement entropy $S_{\!\!f}(L,n)$ takes the form
\begin{equation}
	S_{\!\!f} (L,n)=\frac{N_B}{2} \log_2\frac{n(L-n)}{L} +S_{\!\!f0}.
	\label{slnf}
\end{equation}
On the other hand, for a conformally invariant state, the entanglement entropy $S(L,n)$ takes the form
\begin{equation}
S(L,n)=\frac{c}{3} \log_2 \left(\frac{L}{\pi}\sin{\frac{n\pi}{L}}\right)+S_{\! 0}.
\label{slnfc}
\end{equation}
Note that an additive nonuniversal constant $S_{\!\!f0}$ or $S_{\! 0}$ has been introduced for a scale-invariant or conformally invariant state, respectively.

Three remarks are in order. First, it is necessary to introduce the filling $f$ to indicate distinct degenerate ground states arising from spontaneous symmetry breaking with type-B Goldstone modes, in contrast to  conformally invariant states.
Note that, generically, $f$ refers to a set of fillings $f_1, f_2, \ldots, f_r$, with $r$ being the rank of a symmetry group for a quantum many-body system.
Second, the block does not necessarily consist of contiguous sites for a scale-invariant state, which appears as one of highly degenerate ground states in a quantum many-body system,
as already discussed for the entanglement entropy in the thermodynamic limit~\cite{FMGM}.
Meanwhile, the scaling relation (\ref{slnf}) is left intact as the boundary conditions vary from periodic to open. In contrast,  a conformally invariant state appears to be a non-degenerate ground state for a quantum many-body system at criticality, and the scaling relation (\ref{slnfc}) does depend on the boundary conditions adopted~\cite{cardy,korepin1,zhou}. Third, other choices for the universal scaling function $g(L,n)$ are possible for distinct types of scale-invariant states, but remain to be clarified.

\section{Three illustrative examples}

\subsection{The ${\rm SU}(2)$ spin-$s$ ferromagnetic model}

Consider the ${\rm SU}(2)$ spin-$s$ ferromagnetic Heisenberg model with nearest-neighbor interaction, described by the Hamiltonian
\begin{equation}
\mathscr{H}=-\sum_{j=1}^{L}\textbf{S}_j\cdot \textbf{S}_{j+1}, \label{su2ham}
\end{equation}
where $\textbf{S}_j=(S^x_j,S^y_j,S^z_j)$, and $S^x_j$, $S^y_j$, $S^z_j$ are the spin operators for spin $s$ at the $j$-th site.
Note that for all three illustrative models the sum over $j$ is taken from 1 to $L-1$ for open boundary conditions (OBCs), and from 1 to $L$ for periodic boundary conditions (PBCs).
The spin-$s$ Heisenberg model admits exact solutions as far as its ground state subspace is concerned, though it is only exactly solvable by means of the Bethe ansatz for $s=1/2$.
The degenerate ground states arise from spontaneous symmetry breaking: ${\rm SU}(2) \rightarrow {\rm U}(1)$, thus resulting in one type-B Goldstone mode, so $N_B = 1$.
We remark that the degenerate ground states are identical under both OBCs and PBCs, and span an irreducible representation for the symmetry group ${\rm SU}(2)$, with the dimension being $2sL+1$.

\subsection{The ${\rm SU}(N+1)$ ferromagnetic model}

The ${\rm SU}(N+1)$ ferromagnetic model is described by the Hamiltonian
\begin{equation}
\mathscr{H}=-\sum_{j}P_{j\;j+1}, \label{HsuNp1}
\end{equation}
where $P$ is the permutation operator
$P=\sum_{u,v=1}^{N+1} e_{uv}\otimes e_{vu}$,
where
$e_{uv}=|u\rangle\langle v|$, with $|u\rangle$ and $|v\rangle$ being the $u$-th and $v$-th states in an orthonormal basis.
Physically, the permutation operator $P$ may be realized in terms of the spin-$s$ operators $\textbf{S}=(S_x,S_y,S_z)$, with $N=2s$:
\begin{equation*}
P=\sum_{q=0}^{2s}(-1)^{2s+q} \prod_{m\neq q}^{2s} \frac{2(\textbf{S}\otimes\textbf{S})-m(m+1)+2s(s+1)}{q(q+1)-m(m+1)}.
\end{equation*}
The ${\rm SU}(N+1)$ ferromagnetic model is exactly solvable by means of the nested Bethe ansatz~\cite{sutherland}.
The degenerate ground states arise from spontaneous symmetry breaking: ${\rm SU}(N+1) \rightarrow {\rm SU}(N) \times {\rm U}(1)$ successively,
thus resulting in $N$ type-B Goldstone modes, so $N_B = N$.
We remark that the degenerate ground states  are identical under both OBCs and PBCs, and span an irreducible representation for the symmetry group ${\rm SU}(N+1)$, with the dimension being the combinatorial number $C_{L+N}^N$.

\subsection{The staggered ${\rm SU}(3)$ spin-1  ferromagnetic biquadratic model}

The staggered ${\rm SU}(3)$  spin-1 ferromagnetic biquadratic model is described by the Hamiltonian
\begin{equation}
\mathscr{H}=\sum_{j}{\left(\textbf{S}_j \cdot \textbf{S}_{j+1}\right)^2}. \label{hambq}
\end{equation}
Here $\textbf{S}_j=(S^x_j,S^y_j,S^z_j)$ is the spin-1 operator at site $j$.
The model (\ref{hambq}) is exactly solvable~\cite{barber}. Indeed, it constitutes (up to an additive constant) a representation of the Temperley-Lieb algebra~\cite{tla,baxterbook,martin}, and thus follows from a solution to the Yang-Baxter equation~\cite{baxterbook,sutherlandb,mccoy}.
Note that it is peculiar, in the sense that the ground states are highly degenerate, exponential with the system size $L$, thus leading to non-zero residual entropy~\cite{spins},
in sharp contrast to the ${\rm SU}(2)$ spin-$s$ ferromagnetic model (\ref{su2ham}) and the ${\rm SU}(N+1)$ ferromagnetic model (\ref{HsuNp1}).
Remarkably, the degenerate ground states for this model arise from spontaneous symmetry breaking ${\rm SU}(3) \rightarrow {\rm U}(1) \times {\rm U}(1)$, with the number of the type-B Goldstone modes $N_B=2$, and the ground state degeneracies constitute two Fibonacci-Lucas sequences~\cite{golden} (also cf.~\cite{spins,saleur}) under OBCs and PBCs.

\section{ Universal finite-size scaling for the entanglement entropy $S_{\!\!f}(L,n)$}

For the ${\rm SU}(2)$ spin-$s$ ferromagnetic Heisenberg model, the entanglement entropy $S(L,n,M)$ for the highly degenerate ground states $|L,M\rangle$ have been derived in Ref.~\cite{FMGM}, with $f$ being the filling $f=M/L$. For convenience, the explicit expression for  $S(L,n,M)$ has been collected in Section C of the SM.

\begin{figure}[htb]
	\centering
	\includegraphics[width=0.48\textwidth]{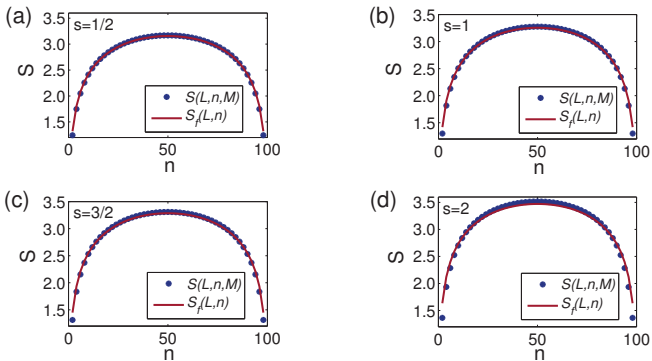}
	\caption{ The entanglement entropy $S(L,n,M)$ vs $n$ against $S_{\!\!f}(L,n)$  vs $n$ for the highly degenerate ground states in the ${\rm SU}(2)$ spin-$s$ ferromagnetic Heisenberg model, with the system size $L=100$, for (a) $s=1/2$, (b) $s=1$, (c) $s=3/2$ and (d) $s=2$.
		The filling factor $f$ is chosen to be $f=1/4$ with in each case $L=100$.
		The best fitting yields (a) $S_{\!\!f0}= 0.834$, (b) $S_{\!\!f0}= 0.934$,  (c) $S_{\!\!f0}= 0.963$, and (d) $S_{\!\!f0}= 1.149$, with the relative errors being less than $1.5\%$,	when $n$ ranges from 10 to 90.}
	\label{comparesu2spins}
\end{figure}

We plot the entanglement entropy $S(L,n,M)$ vs $n$ against the universal finite-size scaling $S_{\!\!f}(L,n)$ vs $n$ in Fig.~\ref{comparesu2spins} for $s=1/2$, $1$, $3/2$ and $2$. Our numerical data for $S(L,n,M)$ fall on the curve $S_{\!\!f}(L,n)$, when $n$ ranges from 10 to 90, with the relative errors being less than $1.5\%$. Here and hereafter, we have regarded
 $S_{\!\!f}(L,n)$ as a function of $n$ for fixed $L$ and $f$.

For the ${\rm SU}(N+1)$ ferromagnetic model, the entanglement entropy $S(L,n,M_1,\ldots,M_N)$ for the highly degenerate ground state $|L,M_1,M_2,\ldots, M_N\rangle$  has been derived in Ref.~\cite{FMGM}, with the filling factors $f_1=M_1/L$, $f_2=M_2/L$ and $f_3=M_3/L$.  The explicit expression for $S(L,n,M_1,\ldots,M_N)$ has been collected in Section C of the SM.

\begin{figure}[htb]
	\centering
	\includegraphics[width=0.48\textwidth]{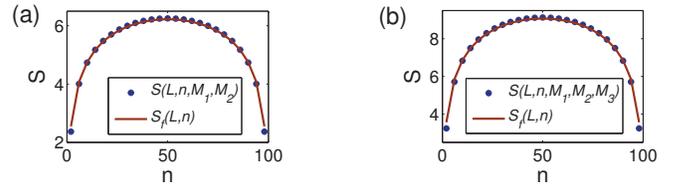}
	\caption{
		(a) The entanglement entropy $S(L,n,M_1,M_2)$ vs $n$ against $S_{\!\!f}(L,n)$ vs $n$ for the highly degenerate ground states in the ${\rm SU}(3)$ spin-$1$ model. The filling factors are chosen to be $f_1=1/4$ and $f_2=1/4$.
		(b) The entanglement entropy $S(L,n,M_1,M_2,M_3)$ vs $n$ against $S_{\!\!f}(L,n)$ vs $n$ for the highly degenerate ground states in the ${\rm SU}(4)$ spin-$3/2$ model.  The filling factors are chosen to be $f_1=1/4$, $f_2=1/4$ and $f_3=1/4$. For both figures the system size $L=100$.
		The best fitting yields (a) $S_{\!f0}= 1.576$ and (b) $S_{\!f0}=2.102$, with the relative errors being less than $1\%$, when $n$ ranges from 10 to 90.}
	\label{comparesu3su4}
\end{figure}

We plot the entanglement entropy $S(L,n,M_1,M_2)$ and the entanglement entropy $S(L,n,M_1,M_2,M_3)$ vs $n$, with the selected fillings (a) $f_1=1/4$ and $f_2=1/4$ and (b) $f_1=1/4$, $f_2=1/4$ and $f_3=1/4$, respectively, against the universal finite-size scaling $S_{\!\!f}(L,n)$ vs $n$, in Fig.~\ref{comparesu3su4}, with $N_B=2$ and $N_B=3$.
Our numerical data for $S(L,n,M_1,M_2)$ and $S(L,n,M_1,M_2,M_3)$ fall on the curve $S_{\!\!f}(L,n)$, when $n$ ranges from 10 to 90, with the relative errors being less than $1\%$.

Finally for the staggered ${\rm SU}(3)$ spin-1  ferromagnetic biquadratic model, the entanglement entropy $S(L,n,M_1,M_2)$ and the entanglement entropy $S(L,n,M_2,M_3)$ for the two  degenerate ground states $|L,M_1,M_2\rangle_2$ and $|L,M_2,M_3\rangle_4$ have been derived in Ref.~\cite{golden}, with the filling factors $f_1=M_1/L$, $f_2=M_2/L$ and $f_3=M_3/L$.
Here the explicit expressions for $S(L,n,M_1,M_2)$ and $S(L,n,M_2,M_3)$ have been collected in  Section C of the SM.

\begin{figure}[htb]
	\centering
	\includegraphics[width=0.48\textwidth]{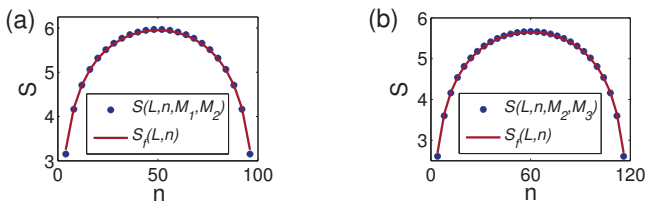}
	\caption{ The entanglement entropy $S(L,n,M_1,M_2)$  vs $n$ against $S_{\!\!f}(L,n)$ vs $n$ for the highly degenerate ground states from the highest weight state in the staggered ${\rm SU}(3)$ spin-1 ferromagnetic biquadratic model, with the system size $L=100$, when the fillings $f_1$ and $f_2$ are chosen to be $f_1=1/4$ and $f_2=1/4$.
		(b) The entanglement entropy $S(L,n,M_2,M_3)$ vs $n$ against $S_f(L,n)$ vs $n$ for the highly degenerate ground states from a generalized highest weight state in the staggered ${\rm SU}(3)$  spin-1 ferromagnetic biquadratic model, with the system size $L=120$, when the fillings $f_2$ and $f_3$ are chosen to be $f_2=1/4$ and $f_3=1/8$.
		The best fitting yields (a) $S_{\!\!f0} =1.308$ and (b) $S_{\!\!f0} =0.743$, respectively, with the relative errors being less than $1\%$,  when $n$ ranges from 10 to $L-10$. }
	\label{comparestsu3}
\end{figure}

For this model we plot the entanglement entropy $S(L,n,M_1,M_2)$ and $S(L,n,M_2,M_3)$ vs $n$ against the universal finite-size scaling $S_{\!\!f}(L,n)$ vs $n$  in Fig.~\ref{comparestsu3}, with the selected filling factors (a) $f_1=1/4$ and $f_2=1/4$, and (b)  $f_2=1/4$ and $f_3=1/8$. Our numerical data for $S(L,n,M_1,M_2)$ and $S(L,n,M_2,M_3)$ are seen to fall on the curve $S_{\!\!f}(L,n)$, when $n$ ranges from 10 to $L-10$, with the relative errors being less than $1\%$.

\section{Summary and outlook}

A universal finite-size scaling behavior of the entanglement entropy has been shown for scale-invariant, but not conformally invariant, degenerate ground states, which arise from spontaneous symmetry breaking with type-B Goldstone modes in quantum many-body systems.
As it turns out, scale-invariant states exhibit a universal finite system-size scaling behavior,  distinct from  conformally invariant states, given both are reduced to a logarithmic scaling function in the thermodynamic limit.
This fact enables us to distinguish  scale-invariant states from conformally invariant states.
It follows that scale invariance does not necessarily imply conformal invariance, in contrast to the speculation made by Polyakov~\cite{polyakov}, even for the familiar ${\rm SU}(2)$ spin-$1/2$ ferromagnetic states -- a paradigmatic example for spontaneous symmetry breaking with type-B Goldstone modes.
Our findings shed further light on a complete classification of quantum phase transitions and quantum states of matter~\cite{entropy}.
In this sense, the possibility for formulating a fully-fledged theory of scale-invariant states remains largely unexplored.

Instead, a simple but heuristic physical argument has been proposed, which imposes three given constraints on the entanglement entropy. This in turn makes it possible to classify distinct types of scale-invariant states. Our results clearly suggest that highly degenerate ground states  arising from spontaneous symmetry breaking with type-B Goldstone modes fall into the simplest category for all possible scale-invariant states, featuring the most fundamental universal scaling function. All other possible universal scaling functions might be expressed in terms of this  fundamental universal scaling function, as illustrated for conformally invariant states.

\section{Acknowledgements}

We thank John Fjaerestad for enlightening discussions.

\newpage
\onecolumngrid
\newpage
\section*{Supplementary Material}
\twocolumngrid
\setcounter{page}{1}
\setcounter{equation}{0}
\setcounter{figure}{0}
\renewcommand{\theequation}{S\arabic{equation}}
\renewcommand{\thefigure}{S\arabic{figure}}

\subsection{A: Derivation of the distinct universal scaling functions $k(x)$ and $k_c(x)$}

As mentioned in the main text, the observation that if a function $k(x)$ yields a solution to the three given constraints, then any function $F(k(x))$ of $k(x)$ also yields a solution, subject to the two conditions for $F(x)$: its constant term vanishes, $F(0)=0$, and its first-order derivative is nonzero, $F'(0) \neq 0$.
Hence, one may search for the most fundamental solution, which is supposed to be a universal scaling function {\it solely} as a result of a scale transformation.

Keeping this fact in mind, we may seek for a universal scaling function $k(x)$ of the form
\begin{equation}
k(x)=u(x)u(1-x).
\label{kux}
\end{equation}
This guarantees that the constraint (3) is satisfied. Here, $u(x)$ is  a polynomial function of $x$, without a constant term,
i.e., $u(x)=\sum _{\beta =1}^m \nu_\beta x^\beta$, with $m$ being an integer, and  $\nu_1, \nu_2,\ldots, \nu_m$ being real numbers yet to be determined.
The constraint (4) yields  $\nu_1 \sum _{\beta =1}^m \nu_\beta=1$. As a consequence, we have to distinguish the two situations: (1) $\nu_1=1$ and (2) $\nu_1 \neq 1$.

If $\nu_1=1$, then we have $\sum _{\beta =2}^m \nu_\beta=0$.
Hence, the simplest choice for $u(x)$ is $u(x)=x$, which amounts to stating that $\nu_1=1$ and $\nu_\beta=0$ ($\beta =2, \ldots, m$). Accordingly, we have
\begin{equation}
k(x)=x(1-x).\label{kx}
\end{equation}
This yields the simplest universal scaling function, which in turn constitutes the most fundamental solution to the three constraints for scale-invariant states.

Actually, this is the {\it only} solution available, as long as  $u(x)$ is  a polynomial function of $x$. A proof for our claim is based on  mathematical induction.
In fact, for $m=1$, obviously we have $\nu_1=1$, and for $m=2$, it is straightforward to see that  $\nu_1=1$ and $\nu_2=0$. Suppose our claim is valid for $m$. Then, for $m+1$, one might resort to
the following lemma: if $u(x)$ yields a solution $k(x)$ to the three constraints, then  $\tilde u(x) =(u(x) - x^2)/(1-x)$ also yields a solution.
Indeed,  the mapping from $u(x)$ to $\tilde u(x)$ turns a polynomial function $u(x)$ of $x$, with order $m+1$, into a polynomial function $\tilde u(x)$ of $x$, with order $m$, due to the fact that $u(x) - x^2$ may be factorized into $1-x$ times a polynomial function of $x$, with order $m$.

If $\nu_1 = t \neq 1$, then we have $\sum _{\beta =2}^m \nu_\beta=t^{-1} - t$, with $t$ being a nonzero real number. In this situation, we have different possible choices for
a universal scaling function $k(x)$, as long as it is monotonically increasing with $x$, when $x$ varies from $x=0$ to $x=1/2$. In fact, it even makes sense to take the limit
$m \rightarrow \infty$, with properly chosen $t$ and $\nu_\beta$. In particular,  if we choose $t= \sqrt{\pi/2}$, $\nu_\beta = (-1)^{\delta} (\pi/2)^{2\delta+1/2}/[(2\delta+1) \times \cdots \times 1]$ for $\beta =2\delta+1$ and
 $\nu_\beta =0$ for $\beta =2\delta$, with $\delta =0,1,2,\ldots$. Then we have $u(x) = \sqrt{2/\pi} \sin(\pi x/2)$. Thus we are led to another universal scaling function $k_c(c)$, namely
\begin{equation}
	k_c(x)=\frac{\sin (\pi x)}{\pi}.
	\label{hc}
\end{equation}
This is precisely the universal scaling function, as predicted by Calabrese and Cardy~\cite{cardy-S}, for conformally invariant states.

Two remarks are in order. First, $k_c(x)$  {\it not only} admits a factorization in a multiplicative way, {\it but also} splits in an additive way. This might be attributed to
the fact that the Lorentz invariance is part of the symmetry group, in addition to a scale transformation, for conformally invariant states.
In fact, $k_c(x)$ may be constructed as follows.
In order to satisfy the constraint (3), we assume that $k_c(x)=v(x)+v(1-x)$, with $v(x)$ being a non-polynomial function of $x$.
As a possible trial solution, we choose $v(x)=b\exp(ax)$, with $a$ and $b$ being real numbers yet to be determined. Hence, we have
\begin{equation}
	k_c(x)=2b\exp(a/2)\cosh[a(x-\frac{1}{2})].
	\label{hab}
\end{equation}
When $x = 0$,  $k_c(x)$ must vanish. Thus, we have $a=i\pi$.
Substituting  $a=i\pi$ back into (\ref{hab}), we have $k_c(x)=2ib\sin(\pi x)$.
To ensure that the constraint (4) is satisfied, we have $k_c(x) \rightarrow x$ for $x\rightarrow 0$. This implies that  $b=1/(2i\pi)$.
Hence, we are led to the same universal scaling function as  that in Eq.~(\ref{hc}).
Second, as follows from Eq.~(\ref{kx}), $x$ may be re-expressed in terms of $k(x)$ as an inverse function, i.e., $x=[1\pm \sqrt{1-4k(x)}]/2$.
Hence,  $k_c(x)$ turns out to be  a function of $k(x)$, with
$k_c(x) = 1/\pi \cos [\pi/2\sqrt{1-4k(x)}]$. This provides an illustrative example for the observation discussed in the main text.

The justification for $k(x)$ as the universal scaling function for  highly degenerate ground states arising from spontaneous symmetry breaking with type-B Goldstone modes
lies in the fact that {\it only} a scale transformation is involved. Further confirmation comes from an asymptotic analysis of
the entanglement entropy for the ${\rm SU}(2)$ spin-$1/2$ ferromagnetic states, as discussed in the next Section.

\subsection{B: Entanglement entropy for the ${\rm SU}(2)$ spin-$1/2$ ferromagnetic states}

Alternatively, the scaling relations of the entanglement entropy $S_{\!\!f}(L,n)$ for both scale-invariant and conformally invariant states, as presented in
Eq.(8)  and Eq.(9) in the main text, respectively, may be justified by extending a heuristic argument in Ref.~\cite{FMGM-S} in the thermodynamic limit.

Given that the number of the type-B Goldstone modes and central charge measure the ability for a system to react when a length scale is present and count the number of low-lying gapless excitations, we anticipate that, for scale-invariant states, $S_f(L,n)$ takes the form
\begin{equation}
S_{\!\!f}(L,n)= \frac{N_B}{2} \log_2 g(L,n)+S_{\!\!f0},
	\label{sf}
\end{equation}
where $g(L,n)=L \, k({n}/{L})$, with $k(n/L)$ being a function of $n/L$, yet to be determined.
This may be determined from  an exactly solvable quantum many-body model -- the ${\rm SU}(2)$ spin-$1/2$ ferromagnetic model, with $N_B=1$.

For the ${\rm SU}(2)$ spin-$1/2$ ferromagnetic states, the entanglement entropy $S(L,n,M)$ for a degenerate ground state $|L,M\rangle$ has been
derived in Refs.~\cite{popkov-S,FMGM-S}. It takes the form
\begin{equation}
S(L,n,M)= -\sum_{k=1}^{n}\Lambda(L,n,k,M)\log_{2}\Lambda(L,n,k,M),\label{s12lnm}
\end{equation}
where the eigenvalues $\Lambda(L,n,k,M)$ of the reduced density matrix $\rho_L(n,M)$ are~\cite{popkov-S,FMGM-S}
\begin{equation}
\Lambda(L,n,k,M)=\frac{C_n^kC_{L-n}^{M-k}}{C_L^M}.
\end{equation}

The eigenvalues of the reduced density matrix,  for large enough $L$, may be approximated as a result of the Stirling approximation: $m!=m^m\exp(-m)\sqrt{2\pi m}$,
 \begin{equation}
\Lambda(L,n,k,M)=C_n^kp^kq^{n-k},
\label{cnkpq}
 \end{equation}
where $p=M/L$ and $q=1-p$.

For large enough $n\gg 1$, the dominant contribution
to the sum (\ref{s12lnm}) comes from the eigenvalues $\Lambda(L,n,k,M)$ with large $k$'s.
Indeed, the binomial coefficients in (\ref{cnkpq}) with $0<p<1$ may be replaced by the normal
distribution (cf. Ref.~\cite{Allen-S}):
\begin{equation}
C_{n}^{m}p^{m}q^{n-m}\approx \frac{1}{\sqrt{2\pi npq}}\exp \left( -\frac{%
	(m-np)^{2}}{2npq}\right) ,\text{ \ \ }npq\gg 1,  \label{binom_approx}
\end{equation}

Hence, the eigenvalues in Eq.~(\ref{cnkpq}) become
\begin{align}
\Lambda (L,n,k,M)& =\frac{%
	C_{k}^{n}p^{k}q^{n-k}C_{M-k}^{L-n}p^{M-k}q^{L-n-M+k}}{C_{M}^{L}p^{M}q^{L-M}} \nonumber
\\
& \approx \frac{1}{n}\frac{1}{\sqrt{2\pi \alpha }}\exp \left( -\frac{(\frac{k%
	}{n}-p)^{2}}{2\alpha }\right) ,
\label{lamn}
\end{align}
where $\alpha =pq(L-n)/nL$.

Substituting (\ref{lamn}) into (\ref{s12lnm}) and replacing the sum with an integral, $S(L,n,M)$ takes the form
\begin{align*}
& S(L,n,M)\approx \int_{0}^{1}R\left( \log _{2}{\frac{R}{n}}\right) dx, \\
& \qquad R=\frac{1}{\sqrt{2\pi \alpha }}\exp \left( -\frac{(x-p)^{2}}{2\alpha }%
\right) .
\end{align*}

For large enough $n$, it yields
\begin{equation}
S(L,n,M)\approx \frac{1}{2}\log _{2}(2\pi epq)+\frac{1}{2}\log
_{2}{\frac{n(L-n)}{L}}. \label{entropy_L}
\end{equation}
We remark that (\ref{entropy_L}) is asymptotically valid for $npq\gg 1$ and becomes exact in
the limit $npq\rightarrow \infty$.

Comparing Eq. (\ref{entropy_L}) with Eq. (\ref{sf}), we are led to conclude that  $k(x) = x(1-x)$.
Hence the scaling relation of the entanglement entropy $S_{\!\!f}(L,n)$ with $L$ and $n$ in Eq.~(8) for scale-invariant states is established.

This argument also works for the scaling relation of the entanglement entropy $S(L,n)$ with $L$ and $n$ in Eq.~(9) for conformally invariant states, since the exact results are available for the transverse-field spin-$1/2$ Ising model at its critical point, with central charge $c=1/2$~\cite{korepin-S}.

~

\subsection{C: Eigenvalues of the reduced density matrices for the highly degenerate ground states}

Here we collect the explicit expressions for the eigenvalues of the reduced density matrices, which have been derived in Refs.~\cite{FMGM-S,golden-S} for the
highly degenerate ground states of the three models under consideration.

For the ${\rm SU}(2)$ spin-$s$ ferromagnetic Heisenberg model, the entanglement entropy $S(L,n,M)$ for a highly degenerate ground state $|L,M\rangle$ has been derived in Ref.~\cite{FMGM-S}. It takes the form
\begin{equation}
S(L,n,M)= -\sum_{k=0}^{2sn}\Lambda(L,n,k,M)\log_{2}\Lambda(L,n,k,M),\label{su2slnf}
\end{equation}
where the eigenvalues $\Lambda(L,n,k,M)$ of the reduced density matrix $\rho_L(n,M)$ are
\begin{equation}
\Lambda(L,n,k,M)=\frac{\mu(L,k,M)}{\nu(L,k,M)},
\end{equation}
with
\begin{equation*}
\mu(L,k,M)\!=\!{\sum}'_{n_{-\!s},...,\; n_{s},\atop l_{-\!s},...,\;l_{s}}\prod_{r,t=-s}^{s-1}\!\varepsilon(s,r)^{n_{r}}
{C_{n\!-\!\sum_{m=-s}^{r-1}\!n_m}^{n_r}}\!\varepsilon(s,t)^{l_{t}}{C_{L\!-\!n-\!\sum_{m=-s}^{t-1}l_m}^{l_t}},
\end{equation*}
and
\begin{equation*}
\nu(L,k,M)={\sum}'_{N_{-s}, ...,N_{s}}\prod_{r=-s}^{s-1}\varepsilon(s,r)^{N_{r}}
{C_{L-\sum_{m=-s}^{r-1}N_m}^{N_r}}.
\end{equation*}
Here, the sum $\sum'_{n_{-s},\ldots,\;n_{s}}$ is taken over all possible values of $n_{-s}$, \ldots , $n_s$, subject to the constraints $\sum_{m=-s}^s n_m=n$ and $\sum_{m=-s}^{s}(s-m)n_m=k$, $\sum'_{l_{-s},\ldots,\;l_{s}}$ is taken over all the possible values of $l_{-s}$, \ldots, $l_s$, subject to the constraints: $\sum_{m=-s}^s l_m=L-n$.
The sum $\sum_{m=-s}^{s}(s-m)l_m=M-k$,   $\sum'_{N_{-s},...,\;N_s}$ is taken over all possible values of $N_{-s}$, \ldots, $N_s$, subject to the constraints $\sum_{m=-s}^s N_m=L$ and $\sum_{m=-s}^{s}(s-m)N_m=M$.
The factor $\varepsilon(s,r)$  takes the form
\begin{equation*}
\varepsilon(s,r)=\frac{\prod_{m=r+1}^{s}{(s+m)(s-m+1)}}{\prod_{m=r}^{s-1}(s-m)^2}.
\end{equation*}

For the ${\rm SU}(N+1)$ ferromagnetic model, the entanglement entropy $S(L,n,M_1,\ldots,M_N)$ for a highly degenerate ground state $|L,M_1,M_2,\ldots, M_N\rangle$  has been derived in Ref.~\cite{FMGM-S}.
It takes the form
\begin{widetext}
	\begin{equation}
	S(L,n,M_1,\ldots,M_N)= -\sum_{k_1,\ldots,k_N=0}^{n}\Lambda(L,n,k_1,\ldots,k_N,M_1,\ldots,M_N)\log_{2}\Lambda(L,n,k_1,\ldots,k_N,M_1,\ldots,M_N),\label{sunslnf}
	\end{equation}
	where the eigenvalues $\Lambda(L,n,k_1,\ldots,k_N,M_1,\ldots,M_N)$ of the reduced density matrix $\rho_L(n,M_1,\ldots,M_N)$ are
	\begin{align}
	\Lambda(L,n,k_1,\ldots,k_N,M_1,\ldots,M_N)=
	\frac{\prod_{\alpha=1}^N {C_{n-\sum_{\beta=1}^{\alpha-1}{k_\beta}}^{k_{\alpha}}\prod_{\gamma=1}^N C_{L-n-\sum_{\beta=1}^{\gamma-1}{(M_\beta-k_\beta)}}^{M_\gamma-k_{\gamma}}}}
	{\prod_{\alpha=1}^N {C_{L-\sum_{\beta=1}^{\alpha-1}{M_\beta}}^{M_{\alpha}}}} \;.
	\end{align}
\end{widetext}

For the staggered ${\rm SU}(3)$ spin-1  ferromagnetic biquadratic model, $S(L,n,M_1,M_2)$ and $S(L,n,M_2,M_3)$ for two highly degenerate ground states $|L,M_1,M_2\rangle_2$ and $|L,M_2,M_3\rangle_4$, with the period being 2 and 4, respectively, have been derived in Ref.~\cite{golden-S}.
The entanglement entropy $S(L,n,M_1,M_2)$ is
\begin{widetext}
	\begin{equation}
	S(L,n,M_1,M_2)= -\sum\limits_{k_1=0}^{\min(M_1,n/2)}\sum\limits_{k_2=0}^{\min(M_2,n-k_1)}\Lambda(L,n,k_1,k_2,M_1,M_2)\log_{2}\Lambda(L,n,k_1,k_2,M_1,M_2),\label{stsu3slnf1f2}
	\end{equation}
where $n$ is a multiple of two and the eigenvalues $\Lambda(L,n,k_1,k_2,M_1,M_2)$ of the reduced density matrix $\rho_L(n,M_1,M_2)$ are
\begin{equation}
\Lambda(L,n,k_1,k_2,M_1,M_2)=\frac{C_{n/2}^{k_1}C_{n-k_1}^{k_2}C_{(L-n)/2}^{M_1-k_1}C_{L-n-M_1+k_1}^{M_2-k_2}} {C_{L/2}^{M_1}C_{L-M_1}^{M_2}}.
\label{lamm1m2}
\end{equation}
	The entanglement entropy $S(L,n,M_2,M_3)$ is
	\begin{equation}
	S(L,n,M_2,M_3)= -\sum\limits_{k_2=0}^{3n/4-k_3}\sum\limits_{k_3=0}^{n/4}\Lambda(L,k_2,k_3,M_2,M_3)\log_{2}\Lambda(L,k_2,k_3,M_2,M_3),\label{stsu3slnf2f3}
	\end{equation}
\end{widetext}
where $n$ is a multiple of four, and the eigenvalues $\Lambda(L,n,k_2,k_3,M_2,M_3)$ are
\begin{equation}
\Lambda(L,n,k_2,k_3,M_2,M_3)=\frac{C_{n/4}^{k_3}C_{3n/4-k_3}^{k_2}C_{(L-n)/4}^{M_3-k_3}C_{3(L-n)/4-M_3+k_3}^{M_2-k_2}} {C_{L/4}^{M_3}C_{3L/4-M_3}^{M_2}}.
\label{lamm2m3}
\end{equation}

\end{document}